# Fast DOA estimation using wavelet denoising on MIMO fading channel

A.V. Meenakshi , V.Punitham, R.Kayalvizhi,  S.Asha

Assistant Professor/ECE, Periyar Maniammai University**,** Thanjavur

meenu_gow@yahoo.com, puniwell@yahoo.co.in, kayal2007@gmail.com , ashasugumar@gmail.com

*Abstract*─ This paper presents a tool for the analysis, and simulation of direction-of-arrival (DOA) estimation in wireless mobile communication systems over the fading channel. It reviews two methods of Direction of arrival (DOA) estimation algorithm. The standard Multiple Signal Classification (MUSIC) can be obtained from the subspace based methods. In improved MUSIC procedure called Cyclic MUSIC, it can automatically classify the signals as desired and undesired based on the known spectral correlation property and estimate only the desired signal's DOA. In this paper, the DOA estimation algorithm using the de-noising pre-processing based on time-frequency conversion analysis was proposed, and the performances were analyzed. This is focused on the improvement of DOA estimation at a lower SNR and interference environment. This paper provides a fairly complete image of the performance and statistical efficiency of each of above two methods with QPSK signal.

## Introduction

The goal of direction-of-arrival (DOA) estimation is to use the data received on the downlink at the base-station sensor array to estimate the directions of the signals from the desired mobile users as well as the directions of interference signals. The results of DOA estimation are then used by to adjust the weights of the adaptive beam former.  So that the radiated power is maximized towards the desired users, and radiation nulls are placed in the directions of interference signals. Hence, a successful design of an adaptive array depends highly on the choice of the DOA Estimation algorithm which should be highly accurate and robust. Array signal processing has found important applications in diverse fields such as Radar, Sonar, Communications and Seismic explorations. The problem of estimating the DOA of narrow band signals using antenna arrays has been analyzed intensively over fast few years.[1]-[9].

The wavelet denoising is a useful tool for various applications of image processing and acoustic signal processing for noise reduction. There are some trials for DOA estimation by applying the wavelet transform method into several sub bands MUSIC and CYCLIC MUSIC scenarios [6{8]. But they do not consider larger noise bandwidth with interference signal included in processing samples. In this paper, the DOA estimation algorithm using a time-frequency conversion pre-processing method with a signal OBW (Occupied Bandwidth) analysis was proposed for CYCLIC MUSIC and the effectiveness was verified through the simulation. This is focused on the improvement of DOA estimation performance at lower SNR and interference environment. This is in compliance with the radio usage trends of lower power and widening signal bandwidth especially.

This paper is organized as follows. Section I presents the narrow band signal model with QPSK signal. In section II the above mentioned data model is extended to multi path fading channel. Here we describe two-channel models namely coherent and non-coherent frequency selective slow fading channels. Section III. a and b briefly describes the algorithms we have used. Section IV deal with MUSIC and Cyclic MUSIC algorithms with proposed model. MUSIC procedures are computationally much simpler than the MLM but they provide less accurate estimate [2]. The popular methods of Direction finding such as MUSIC suffer from various drawbacks such as 1.The total number of signals impinges on the antenna array is less than the total number of receiving antenna array. 2. Inability to resolve the closely spaced signals 3. Need for the knowledge of the existing characteristics such as noise characteristics. Cyclic MUSIC algorithm overcomes the above drawbacks. It exhibits cyclostationarity, which improves the DOA estimation. Finally Section V describes the simulation results and performance comparison. Section VI concludes the paper.

---

A.V. Meenakshi is with the Periyar Maniammai University-Thanjavur
e-mail : meenu_gow@yahoo.com

## I. NARROW BAND SIGNAL MODEL

The algorithm starts by constructing a real-life signal model. Consider a number of plane waves from $M$ narrow-band sources impinging from different angles $\theta_i$, $i = 1, 2, \ldots, M$, impinging into a uniform linear array (ULA) of $N$ equi-spaced sensors, as shown in Figure 1.

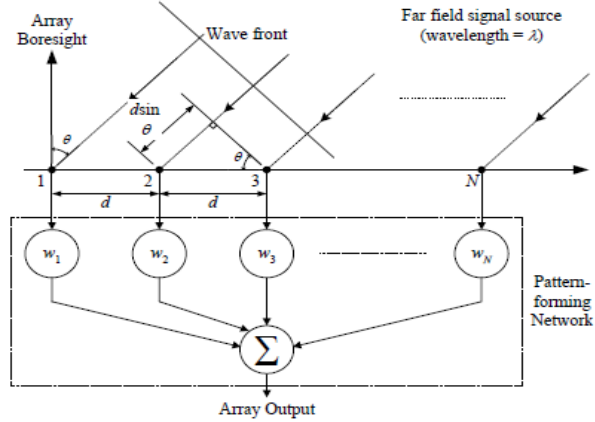

**Figure1.Uniform linear array antenna**

In narrowband array processing, when n signals arrive at an m-element array, the linear data model

$$y(t) = A(\Phi)x(t) + n(t) \qquad (1)$$

is commonly used, where the m*n spatial matrix $A=[a_1, a_2, \ldots, a_n]$ represents the mixing matrix or the steering matrix. In direction finding problems, we require A to have a known structure, and each column of A corresponds to a single arrival and carries a clear bearing. $\mathbf{a}(\Phi)$ is an $N \times 1$ vector referred to as the array response to that source or array steering vector for that direction. It is given by:

$$\mathbf{a}(\Phi) = [1 \; e^{-j\phi} \; \ldots \; e^{-j(N-1)\phi}]^T \qquad (2)$$

where $T$ is the transposition operator, and $\phi$ represents the electrical phase shift from element to element along the array. This can be defined by:

$$\phi = (2\pi/\lambda) d \cos\theta \qquad (3)$$

where $d$ is the element spacing and $\lambda$ is the wavelength of the received signal.

Due to the mixture of the signals at each antenna, the elements of the m×1 data vector y(t) are multicomponent signals. Whereas each source signal x(t) of the n×1 signal vector, x(t) is often a monocomponent signal. n(t) is an additive noise vector whose elements are modeled as stationary, spatially and temporally white, zero mean complex random processes that are independent of the source signals. That is

$$E[n(t+\tau) n^H(t)] = \sigma\delta(\tau)I$$

$$E[n(t+\tau) n^T(t)] = 0, \quad \text{for any } \tau \qquad (4)$$

Where $\delta(\tau)$ is the delta function, I denotes the identity matrix, $\sigma$ is the noise power at each antenna element, superscripts H and T, respectively, denote conjugate transpose and transpose and E(.) is the statistical expectation operator.

In (1), it is assumed that the number of receiving antenna element is larger than the number of sources, i.e., m>n. Further, matrix A is full column rank, which implies that the steering vectors corresponding to n different angles of arrival are linearly independent. We further assume that the correlation matrix

$$R_{yy} = E[y(t)y^H(t)] \qquad (5)$$

is nonsingular and that the observation period consists of N snapshots with N>m.

Under the above assumptions, the correlation matrix is given by

$$R_{yy} = E[y(t)y^H(t)] = AR_{xx}A^H + \sigma I \qquad (6)$$

Where $R_{xx} = E[(x(t)x^H(t))]$ is the source correlation matrix.

Let $\lambda_1 > \lambda_2 > \lambda_3 \ldots \ldots \lambda_n = \lambda_{n+1} = \ldots \lambda_m = \sigma$ denote the eigen values of $R_{yy}$. It is assumed that $\Lambda_i$, $i=1, 2, 3\ldots\ldots n$ are distinct. The unit norm Eigen vectors associated with the columns of matrix $S = [s_1\ s_2\ \ldots s_n]$, and those corresponding to $\lambda_{n+1} \ldots \lambda_m$ make up matrix $G = [g_1 \ldots\ldots g_{m-n}]$. Since the columns of matrix **A** and **S** span the same subspace, then $A^H G = 0$;

In practice $R_{yy}$ is unknown and, therefore, should be estimated from the available data samples $y(i)$, $i = 1\ 2\ 3\ldots\ldots N$. The estimated correlation matrix is given by

$$R_{yy} = 1/N \sum_{n=1}^{N} (y(t) y^H(t)) \qquad (7)$$

Let $\{s_1, s_2, \ldots\ldots s_n, \ldots\ldots g_{m-n}\}$ denote the unit norm eigen vectors of $R_{yy}$ that are arranged in descending order of the associated eigen values respectively. The statistical properties of the eigen vectors of the sample covariance matrix $R_{yy}$ for the signals modeled as independent processes with additive white Gaussian noise are given in [9].

## II. MIMO SIGNAL MODEL

The MIMO received signal data model is given by

$$y_1(t) = \sum_{k=1}^{K} \alpha_1(k) x_{mk}(t) + n_1(t) \qquad (8)$$

Where $\alpha_1(k) = \alpha(k) a_k(\Phi)$; $a_k(\Phi)$ is the antenna response vector. Where $x_{mk}(t)$ is the signal transmitted by $k^{th}$ user of $m^{th}$ signal, $\alpha_1(k)$ is the fading coefficient for the path connecting user k to the $l^{th}$ antenna, $n_l(t)$ is circularly symmetric complex Gaussian noise. Here we examine two basic channel models [4]. In the first case, fading process for each user is assumed to be constant across the face of the antenna array and we can associate a DOA to the signal. This is called coherent wave front fading. In coherent wave front fading channel the fading parameters for each user is modeled as $\alpha_1(k) = \alpha(k) a_k(\Phi)$, where $\alpha(k)$ is a constant complex fading parameter across the array, $\Phi_k$ is the DOA of the $k^{th}$ user's signal relative to the array geometry, and $a_k(\Phi)$ is the response of the $l^{th}$ antenna element to a narrow band signal arriving from $\Phi_k$. The signal model is represented in vector form as

$$y_1 = \sum_{k=1}^{K} \alpha_1(k) g_{mk}(k) + n_1 \qquad (9)$$

Here $g_{mk}$ is a vector containing the $k^{th}$ user's $m_k^{th}$ signal.

The second model we consider is non-coherent element-to-element fading channel on which each antenna receives a copy of the transmitted signal with a different fading parameter. In this case, the dependency of the array response on the DOA for each user cannot be separated from the fading process, so that no DOA can be exploited for conventional beam forming.

## III. ALGORITHMS

### A. MUSIC

MUSIC is a method for estimating the individual frequencies of multiple times – harmonic signals. MUSIC is now applied to estimate the arrival angle of the particular user [1],[2].

The structure of the exact covariance matrix with the spatial white noise assumption implies that its spectral decomposition is expressed as

$$R = APA^H = U_s \Lambda U^H_s + \sigma^2 U_n U^H_n \qquad (10)$$

Where assuming $APA^H$ to be the full rank, the diagonal matrix Us contains the M largest Eigen values. Since the Eigen vectors in Un (the noise Eigen vectors) are orthogonal to A.

$$U_n a(\phi) = 0, \text{ where } \phi \in \{\phi_1, \phi_2, \ldots \phi_m\} \qquad (11)$$

To allow for unique DOA estimates, the array is usually assumed to be unambiguous; that is, any collection of N steering vectors corresponding to distinct DOAs $\Phi_m$ forms a linearly independent set $\{a_{\Phi_1}, \ldots a_{\Phi_m}\}$. If $a(.)$ satisfies these conditions and P has full rank, then $APA^H$ is also full rank. The above equation is very helpful to locate the DOAs in accurate manner.

Let $\{s_1 \ldots s_n, g_1 \ldots g_{m-n}\}$ denote a unit norm eigenvectors of R, arranged in the descending order of the associated Eigen values, and let Š and Ĝ denote the matrices S and G made of $\{s_I\}$ and $\{g_I\}$ respectively. The Eigen vectors are separated in to the signal and noise Eigen vectors. The orthogonal projector onto the noise subspace is estimated. And the MUSIC 'spatial spectrum' is then defined as

$$f(\phi) = \left[a^*(\phi) \hat{G} \hat{G}^* a(\phi)\right] \qquad (12)$$

$$f(\phi) = \left[a^*(\phi)\left[I - SS^*\right]a(\phi)\right] \quad (13)$$

The MUSIC estimates of $\{\Phi_i\}$ are obtained by picking the n values of $\Phi$ for which $f(\Phi)$ is minimized.

To conclude, for uncorrelated signals, MUSIC estimator has an excellent performance for reasonably large values of N, m and SNR. If the signals are highly correlated, then the MUSIC estimator may be very inefficient even for large values of N, m, and SNR.

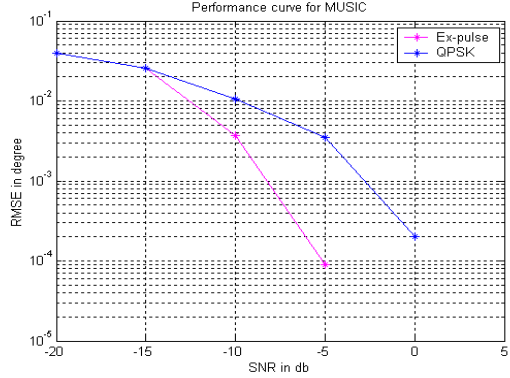

**Figure 2: Performance comparison of MUSIC**

### B. Cyclic MUSIC

We assume that $m_\alpha$ sources emit cyclostationary signals with cycle frequency $\alpha$ ($m_\alpha \leq m$). In the following, we consider that x(t) contains only the $m_\alpha$ signals that exhibit cycle frequency $\alpha$, and all of the remaining $m - m_\alpha$ signals that have not the cycle frequency $\alpha$.

Cyclic autocorrelation matrix and cyclic conjugate autocorrelation matrix at cycle frequency $\alpha$ for some lag parameter $\tau$ are then nonzero and can be estimated by

$$R_{yy}\alpha(\tau) = \sum_{n=1}^{N} y(t_n + \tau/2) y^H(t_n - \tau/2) e^{-j2\pi\alpha t_n} \quad (14)$$

$$R_{yy^*}\alpha(\tau) = \sum_{n=1}^{N} y(t_n + \tau/2) y^T(t_n - \tau/2) e^{-j2\pi\alpha t_n} \quad (15)$$

where N is the number of samples.

Contrary to the covariance matrix exploited by the MUSIC algorithm [1], the Cyclic MUSIC method [8] is generally not hermitian. Then, instead of using the Eigen Value decomposition (EVD), Cyclic MUSIC uses the Singular value decomposition (SVD) of the cyclic correlation matrix. For finite number of time samples, the algorithm can be implemented as follows:

- Estimate the matrix $R_{yy}^\alpha(\tau)$ by using (15) or $R_{yy^*}^\alpha(\tau)$ by using (16).
- Compute SVD

$$\begin{bmatrix} U_s & U_n \end{bmatrix} \begin{bmatrix} \Sigma_s & 0 \\ 0 & \Sigma_n \end{bmatrix} \begin{bmatrix} V_s & V_n \end{bmatrix}^H \quad (16)$$

Where [Us  Un ] and [ Vs  Vn] are unitary, and the diagonal elements of the diagonal matrices $\Sigma_s$ and $\Sigma_n$ are arranged in the decreasing order. $\Sigma_n$ tends to zero as the number of time samples becomes large.

- Find the minima of $\|U_n^H a(\Phi)\|^2$ or the max of $\|U_s^H a(\Phi)\|^2$

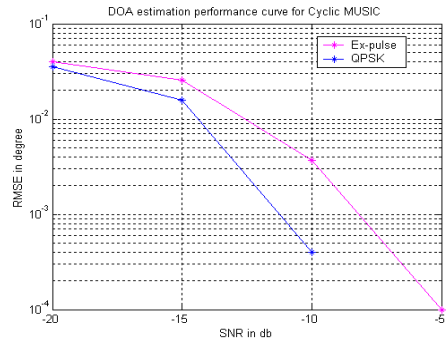

**Figure 3: Performance comparison of Cyclic MUSIC**

## IV. NEW APPROACH OF DOA ESTIMATION

A signal subspace based DOA estimation performance is affected by the two factors of an accurate array manifold modeling and a spatial covariance matrix of a received array signal. A higher SNR signal for a target source is required for an accurate estimation from finite received signal samples. But the DOA estimation performance is limited by the lower SNR from interference signals and environmental noise. For the performance improvement of DOA estimation, this paper proposed a pre-processing technique of time-frequency conversion methodology for signal filtering. This method includes a time-frequency conversion technique with a signal OBW (Occupied Bandwidth) measurement based on wavelet de-noising method as shown in Fig. 1. This is a DOA method for SNR improvement based on time-frequency conversion approach. The improvement of a DOA estimation performance was verified by the simulation.

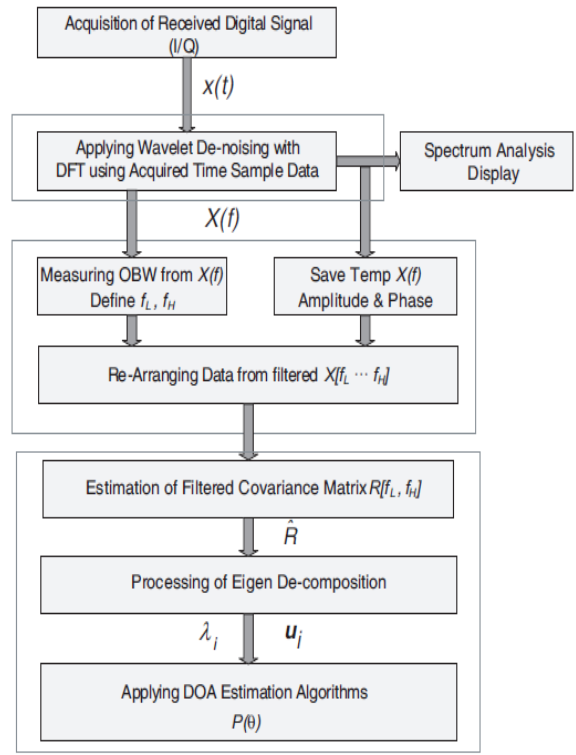

**Figure4.Proposed Model for MUSIC**

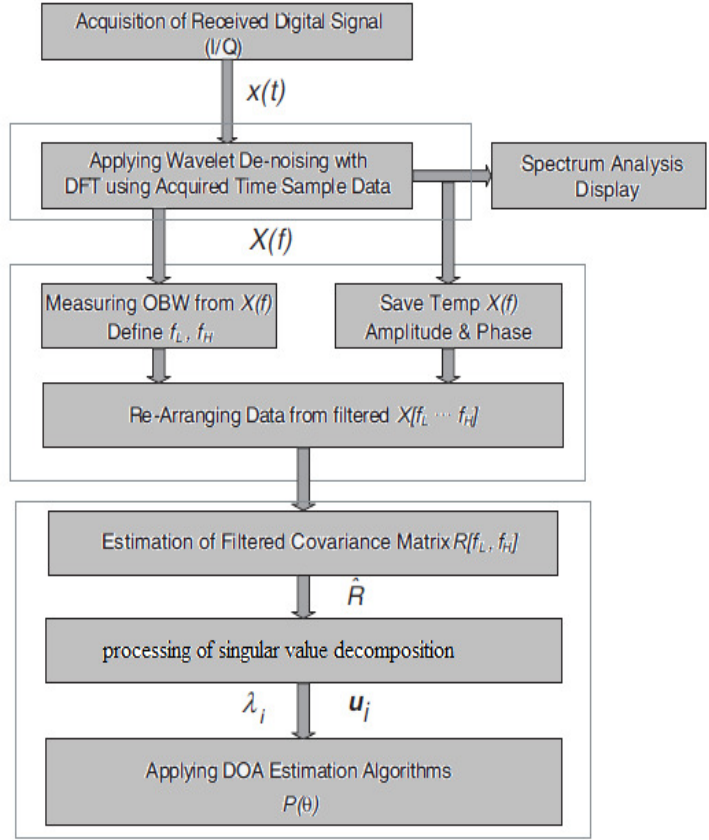

**Figure5:.Proposed Model for CYCLIC MUSIC**

This is proposed to overcome the limitation of existing DOA estimation techniques based on only time domain analysis. The more effective estimation is expected by the improvement of SNR from the proposed pre-processing techniques of frequency domain analysis. The proposed method collects a time sampled signal y(t) from an array antenna as shown in Fig. 1. The upper and lower 99% - OBW limits *fL* and *fH* of a signal are determined from y(*f*), which is the DFT result of a received signal y(*t*). The filtered covariance matrix R[*fL; fH*] can be obtained from the estimated signal energy, y[*fL; fL+1; : : : ; fH*] with an improved SNR. And the more exact OBW measurement is expected through the proposed wavelet de-noising method based on time-frequency analysis. The proposed OBW limits are defined as following measurement concepts of Equations (9) ~ (12). This process can effectively eliminate small interference noises from the target signal streams by the frequency domain analysis [15, 16]. Where *Px* is a power of each spectrum frequency elements { *f1,….., fN*}. The 99% OBW is calculated from the upper limit *fH* and the lower limit *fL* of 0.5% OBW point from each spectrum boundary.

$$p_{rel} = \sum_{i=f_1}^{f_N} Py(i) \qquad (17)$$

$$\Delta P_{\beta/2} = P_{rel} \times \beta/2[\%] \qquad (18)$$

($\beta$ = 1 for 99% OBW analysis)

$$f_L = \arg\min_{f_L} \left\| \sum_{f_L=f_1}^{f_N} Py(f_L) - \Delta P_{\beta/2} \right\| \qquad (19)$$

$$f_H = \arg\min_{f_H} \left\| \sum_{f_H=f_1}^{f_N} Py(f_H) - \Delta P_{\beta/2} \right\| \qquad (20)$$

An improved DOA estimation is expected from the filtered covariance matrix and Eigen-decomposition processing at particularly low SNR signal conditions. By the proposed pre-processing, it can effectively reject adjacent interferences at low SNR conditions. Moreover, it can acquire the signal spectrum with an improved DOA estimation spectrum simultaneously without additional computation. This improved signal spectrum is important results for radio surveillance procedure. The signal de-noising is achieved by the discrete wavelet transform-based thresholding to the resulting coefficients, and suppressing those coefficients smaller than certain amplitude. An appropriate transform can project a signal to the domain where the signal energy is concentrated in a small number of coefficients. The proposed Wavelet de-noising process get a de-noised version of input signal obtained by thresholding the wavelet coefficients. In this paper, the wavelet procedure applied the heuristic soft thresholding of wavelet decomposition at level one. This de-noising processing model is depicted as following simple model.

$$S(n) = f(n) + \sigma e(n), \quad n=0,\ldots,N-1 \qquad (21)$$

In this simplest model, $e(n)$ is a Gaussian white noise of independent normal random variable $N(0; 1)$ and the noise level is supposed to be equal to 1. Using this model, it follows the objectives of noise removal by reconstruct the original signal $f$. It can be assumed that the higher coefficients are result from the signal and the lower coefficients are result from the noise. The noise eliminated signal is obtained by transforming back into the original time domain using these wavelet coefficients.

## V. SIMULATION AND PERFORMANCE COMPARISON

**Data Specification**

Signal specification:
Data Model:   QPSK
    Input bit duration   T   = 0.5μsec
    Sampling interval   t   = T/10;

Antenna Array Model:

Type:  Uniform Linear array antenna
    No. of array Elements   N   = 16
    Free space velocity   c   = $3*10^8$
    Centre frequency   fc   = 2.4GHz
    Wavelength   λ   = c / fc
    Inter element Spacing   d   = λ/2
    Angle of arrival in degrees   θ   = -5 to 20

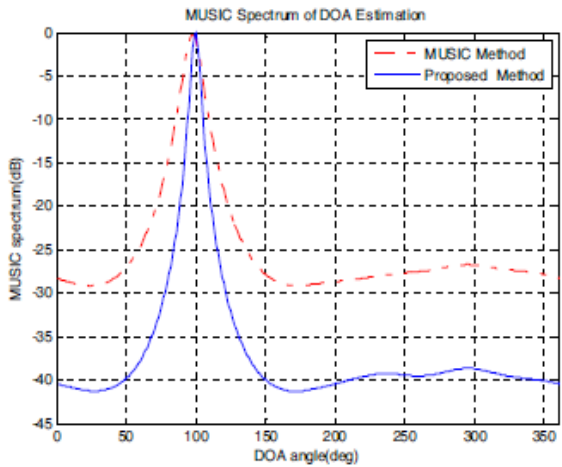

**Figure6. DOA estimation spectrum**

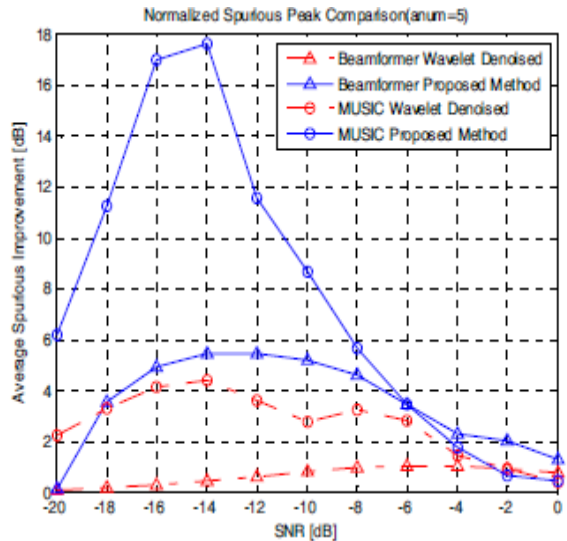

**Figure7. Comparison of spurious peak**

In this section, we present some simulation results, to show the behavior of the three methods and to compare the performance with the analytically obtained Root Mean Squared Error (RMSE). We consider here a linear uniformly spaced array with 16-antenna elements spaced $\lambda/2$ apart. Incoming QPSK signals are generated with additive white Gaussian noise with signal to noise ratio 10dB, the bit rate of the QPSK signal of interest is 2Mb/s and other QPSK modulated signals with data rate 1Mb/s are considered as interference. MUSIC is simulated using the specified parameters. The Cyclic MUSIC algorithm is also simulated with some cyclic frequency of 4MHz and some lag parameter of 2. One QPSK signals arrived at 20 degree and an interferer at 5 degree DOA. RMSE Vs SNR plots for the two methods as shown in figure 2and 3. This section presents Mont Carlo computer simulation results to illustrate the performance of the proposed algorithms for synchronous system. Each Monte Carlo experiment involved a total of 1000 runs, and each estimation algorithm is presented with exactly the same data intern. It is interesting to note that QPSK signal performs better than FM signal. So that bandwidth requirement is as low as possible for QPSK signals as FM signals. It is interesting to note that the conventional MUSIC would require more data samples than Cyclic MUSIC to achieve the same RMSE. Form the simulation results, the proposed method improves the DOA estimation performance of accuracy and spurious peak of spatial spectrum especially for lower SNR signals. Figure 6 and 7 show the comparison results of DOA estimation performance for low SNR signals with interference and noise. The DOA estimation performance was compared by the spurious peak of DOA estimation spectrums which increase a measurement ambiguity and probability of successful DOA estimation. The proposed method improves the spurious peak characteristic more than 10 dB at less than -10 dB SNR signal condition by applying MUSIC and cyclic MUSIC estimation.

## VI. CONCLUSION

Unlike MUSIC, Cyclic MUSIC does not suffer from the drawback of requiring a higher number of antenna elements than sources. Good signal selective capability and high resolution is achieved in Cyclic MUSIC than MUSIC. This algorithm exploits cyclostationarity, which improves the signal Direction of Arrival estimation over the fading channel. The proposed method shows an improved ability of DOA resolution and estimation error at the noise and interference conditions. These are the measurement limits at on-air environment.